\documentclass[%
 reprint,
superscriptaddress,
 amsmath,amssymb,
 aps,
]{revtex4-1}

\usepackage{graphicx}% Include figure files
\usepackage{dcolumn}% Align table columns on decimal point
\usepackage{bm}% bold math
\usepackage{hyperref}% add hypertext capabilities
\begin{document}

\preprint{APS/123-QED}

\title{Test the Weak Cosmic Supervision Conjecture in Dark Matter-Black Hole System
}%

\author{Liping Meng}
% \altaffiliation{College of Physics, Guizhou University, Guiyang 550025, China}%Lines break automatically or can be forced with \\
\author{Zhaoyi Xu}
 \email{Electronic address: zyxu@gzu.edu.cn(Corresponding author)
}
 \author{Meirong Tang}
 \email{Electronic address: tangmr@gzu.edu.cn(Corresponding author)
}
\affiliation{%
College of Physics, Guizhou University,\\
 Guiyang 550025, China
}%





\date{\today}% It is always \today, today,

\begin{abstract}
There is a possibility that the event horizon of a Kerr-like black hole with perfect fluid dark matter (DM) can be destroyed, providing a potential opportunity for understanding the weak cosmic censorship conjecture of black holes. In this study, we analyze the influence of the strength parameter of perfect fluid DM on the destruction of the event horizon of a Kerr-like black hole with spinning after injecting a test particle and a scalar field. We find that, when a test particle is incident on the black hole, the event horizon is destroyed by perfect fluid dark matter for extremal black holes. For nearly extremal black holes, when the dark matter parameter satisfies $\alpha \in \left (-r_{h} , 0\right ) \cup \left ( r_{h} ,k_2\right )$ i.e.$(A<0)$, the event horizon of the black hole will not be destroyed; when the dark matter parameter satisfies $\alpha \in\left ( k_1 ,-r_{h} \right ]\cup \left[0,r_{h}\right ]$ i.e.$(A\ge 0)$, the event horizon of the black hole will be destroyed. When a classical scalar field is incident into the black hole in the extremal black hole case, we find that the range of mode patterns of the scalar field that can disrupt the black hole event horizon is different for different values of the perfect fluid dark matter strength parameter. In the nearly extremal black hole case, through our analysis, we have found when $\alpha\neq0 $ and $\alpha\neq\pm\ r_h$ i.e.$A\neq0$, the event horizon of the black hole can be disrupted. Our research results indicate that dark matter might be capable of breaking the black hole horizon, thus potentially violating the weak cosmic censorship conjecture.
\begin{description}
\item[Keywords]
Dark matter, Kerr black hole, Event horizon, Weak cosmic censorship conjecture, Scalar field
\end{description}
\end{abstract}

\maketitle

%\tableofcontents
\section{Introduction}
The study of dark matter (DM) is currently a hot topic of interest among physicists\cite{Bi2018,Zhou2015}, requiring the integration of cosmology and particle physics. Current observational evidence indicates that DM is nearly invisible and only interacts through gravity. Since DM cannot be directly observed, its existence can only be inferred through indirect evidence\cite{Oort1932TheFE}, such as measurements of galaxy rotation curves\cite{Rubin1980Rotational,Karukes2016TheUR,7a}, large-scale structure measurements\cite{fang1984dark,gao2020large}, weak gravitational lensing effects\cite{zhang2014precise,Weinberg2002}, galaxy mass-to-light ratios\cite{zhong2018}, microwave background radiation measurements\cite{shen2022}, and cluster dynamics measurements\cite{Tagliaferro2021DynamicalAO,1933AcHPh}. The cold dark matter model is one of the predominant models for dark matter\cite{Navarro_1996,Navarro_1997}, which argues that the vast majority of DM is composed of neutral weakly interacting heavy particles\cite{bi2011}, and the number of DM is very large\cite{2014}. It is estimated that DM constitutes approximately 23$\%$ of the total matter content in the present-day universe. Classic examples of evidence include Fritz Zwicky's\cite{1933AcHPh} early suggestion of the existence of dark matter, and Vera Rubin's \cite{Rubin1980Rotational}observations of the motion of stars within galaxies in the 1980s, which revealed a more concentrated distribution of matter in galaxies than predicted by gravitational effects, providing strong support for the theory of dark matter. In 2003, the Wilkinson Microwave Anisotropy Probe (WMAP) produced the first image of the infant universe and accurately measured cosmological parameters. That same year, the Sloan Digital Sky Survey (SDSS) also obtained similar results. The achievements of WMAP and SDSS provide compelling evidence for the existence of dark matter\cite{Charles2003}.

The concept of black holes was initially proposed by J. Michell and P. S. Laplace\cite{Michell1784}, and the term "black hole" was coined by John Wheeler in 1967. Black holes are the result of stellar evolution and interactions, characterized by an extremely strong gravitational field and curvature. In recent years, with the rapid development of gravitational wave astronomy \cite{hu2019,zhu2016,liu2016,abbott2016,abbott2017,abbott2019}and the successful capture of black hole images in 2019\cite{akiyama2019,akiyama2019vi,akiyama2019v}, black holes have become a hot topic of discussion. In the central regions of galaxies, along with a significant amount of dark matter particles, there exist supermassive black holes\cite{lyndenbell1969,eckart1997,Gebhardt_2011,Kormendy1995InwardBS}. Dark matter and black holes can form a stable system in these regions\cite{Batic2022}. Connecting dark matter and black holes in the cores of galaxies allows for a deeper understanding by studying their interactions and dynamics, providing insights into the nature of both dark matter and black holes.

The gravitational collapse inevitably leads to spacetime singularities, as described by the famous Hawking-Penrose singularity theorems\cite{penrose1965,hawking1970}. The existence of spacetime singularities implies the breakdown of gravitational theory. In order to preserve the predictability of gravitational theory, Roger Penrose proposed the weak cosmic censorship conjecture in 1969\cite{penrose1969,carroll2019}. This conjecture relates to the formation process of black holes and suggests that black holes formed in nature satisfy a principle of cosmic censorship, ensuring the existence of what is known as the "weak cosmic censorship conjecture." Under this conjecture, spacetime singularities are hidden behind the event horizons of black holes and are forever invisible to distant observers.

The strict formulation of the weak cosmic censorship conjecture within the framework of general relativity states that any formation of black holes caused by matter must satisfy the requirements of the conjecture, which means that naked singularities do not exist in the universe. Specifically, it requires that every black hole must be enveloped by an event horizon, ensuring that naked singularities are not exposed to the universe. In other words, spacetime singularities resulting from gravitational collapse must necessarily exist within the event horizon. In essence, while the weak cosmic censorship conjecture has not been fully proven and lacks a rigorous mathematical definition, it has become one of the foundations of black hole physics research. The latest research extends the cosmic censorship conjecture to Anti-de Sitter spacetime\cite{gwak2017,Gwak2019WeakCC}.
 
Currently, the weak cosmic censorship conjecture has been validated through various means. For example, numerical evolutions of collapses of dust clouds or other matter fields\cite{christodoulou1984,east2019,song2021,Ames2023OnTH}, extensive nonlinear numerical simulations of perturbed black holes or black hole rings\cite{eperon2020,figueras2017,figueras2016,lehner2010}, and numerical evolutions of black hole collisions and mergers in four dimensions and higher-dimensional spacetimes\cite{sperhake2009,andrade2019,Andrade2020EvidenceFV,Figueras2022EndpointOT,Bantilan2019EndPO}. In the context of dark matter-black hole systems, this study employs thought experiments to investigate how dark matter can modify the weak cosmic censorship conjecture, thus providing potential avenues for studying the interiors of black holes.
 
The article is structured as follows. In Section II, we introduce the relevant basic properties of the Kerr-like black hole with perfect fluid dark matter (PFDM). In Sections III and IV, based on the work in Section II, we investigate the weak cosmic censorship conjecture of the black hole by using a scalar field and a test particle with high angular momentum injected into the Kerr-like black hole with perfect fluid DM. In the last section, we provide a brief summary of the preceding research and discuss the issues addressed in this study.

\section{a Kerr-like black hole with perfect fluid DM}
A Kerr-like black hole with perfect fluid dark matter is a four-dimensional rotating black hole. In Boyer-Lindquist coordinates, the metric of a Kerr-like black hole with perfect fluid dark matter is given by the following equation\cite{xu2018}
\begin{equation}\label{1}
\begin{split}
\mathrm{d}\mathrm{s}^2=&\left(1-\frac{2Mr-\alpha r\ln{(}\frac{r}{\left|\alpha\right|})}{\Sigma^2}\right)dt^2+\frac{\Sigma^2}{\Delta_r}dr^2-\\
&\frac{2a\sin^2{\theta}\left(2Mr-\alpha r\ln{(}\frac{r}{\left|\alpha\right|}\right)}{\Sigma^2}d\varphi dt+\Sigma^2d\theta^2\\
&+{sin}^2{\theta}\left(r^2+a^2+a^2{sin}^2{\theta}\frac{2Mr-\alpha r\ l\ n{(}\frac{r}{\left|\alpha\right|})}{{\sum}^2}\right)d\varphi^2,
\end{split}
\end{equation}
the metric functions $\mathrm{\Sigma}^2$ and $\mathrm{\Delta}_r$ are expressed as follows
\begin{equation}\label{2}
\mathrm{\Sigma}^2=r^2+a^2{cos}^2{\theta},
\end{equation}
\begin{equation}\label{3}
\mathrm{\Delta}_r=r^2-2Mr+a^2+\alpha\ rln{(}\frac{r}{\left|\alpha\right|}).
\end{equation}
Where $M$ represents the black hole mass, $\alpha$ represents a parameter describing the strength of perfect fluid dark matter, $a=\frac{J}{M}$ represents the black hole spin parameter, and $J=Ma$ represents the angular momentum of the black hole. If the PFDM does not exist, i.e.$\alpha=0$, the metric mentioned above will become the metric of a Kerr black hole.

Under the metric, the event horizon $r_h$ of a Kerr-like black hole with perfect fluid dark matter is defined by equation $g^{rr}=0$ i.e.$\mathrm{\Delta}_r=0$. Here, we set a parameter 
\begin{equation}\label{4}
A=\alpha\ r_hln{(}\frac{r_h}{\left|\alpha\right|}),
\end{equation}

\begin{figure}[h]
\includegraphics[width=0.5\textwidth]{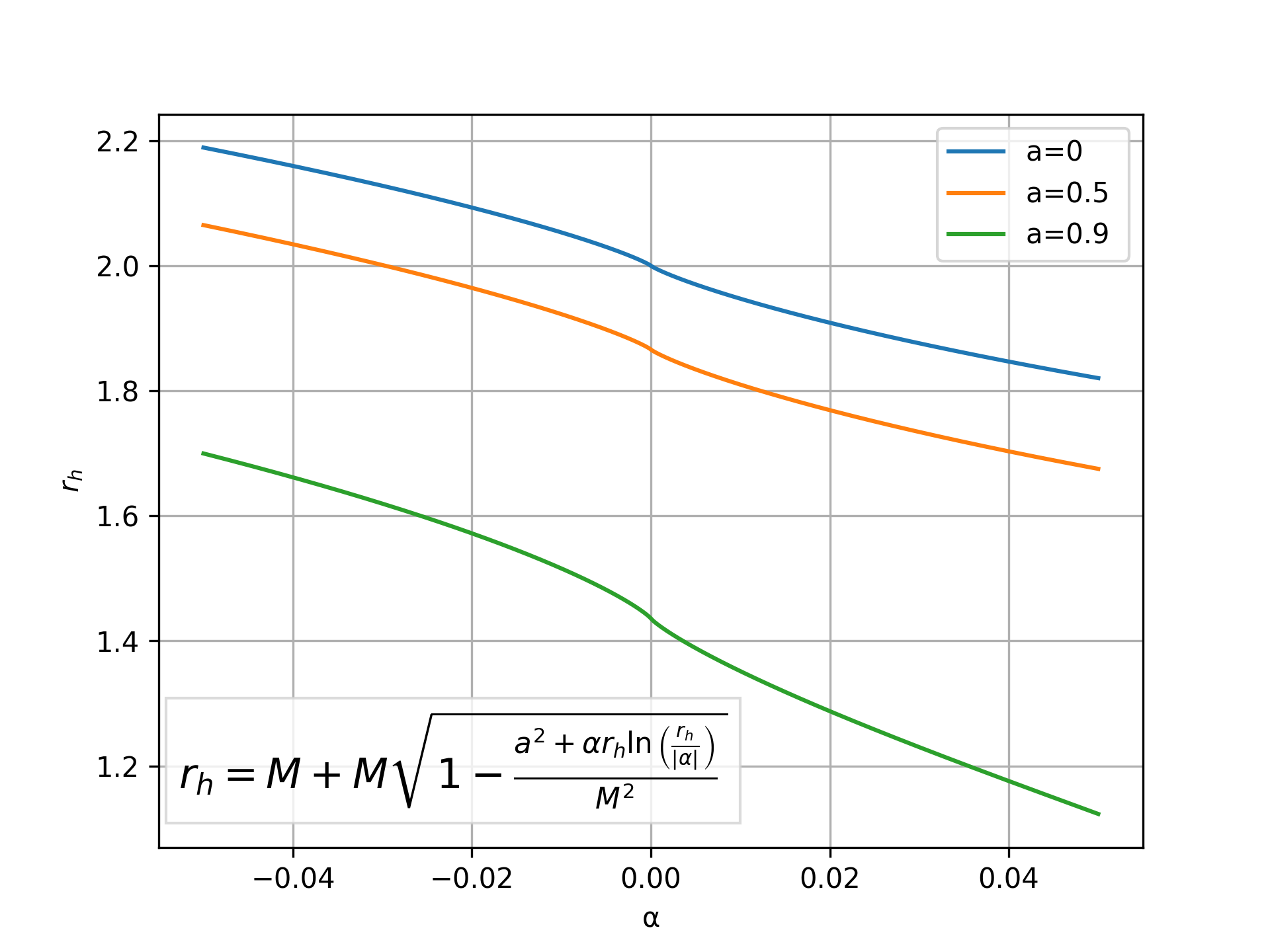}% Here is how to import EPS art
\caption{This is a plot showing the relationship between the event horizon $r_h$ and the strength parameter $\alpha$ of the perfect fluid DM, where the black hole mass $M=1$, the spin parameter $a=0,0.5,0.9$.}
\label{a}
\end{figure}

\begin{figure}[h]
\includegraphics[width=0.5\textwidth]{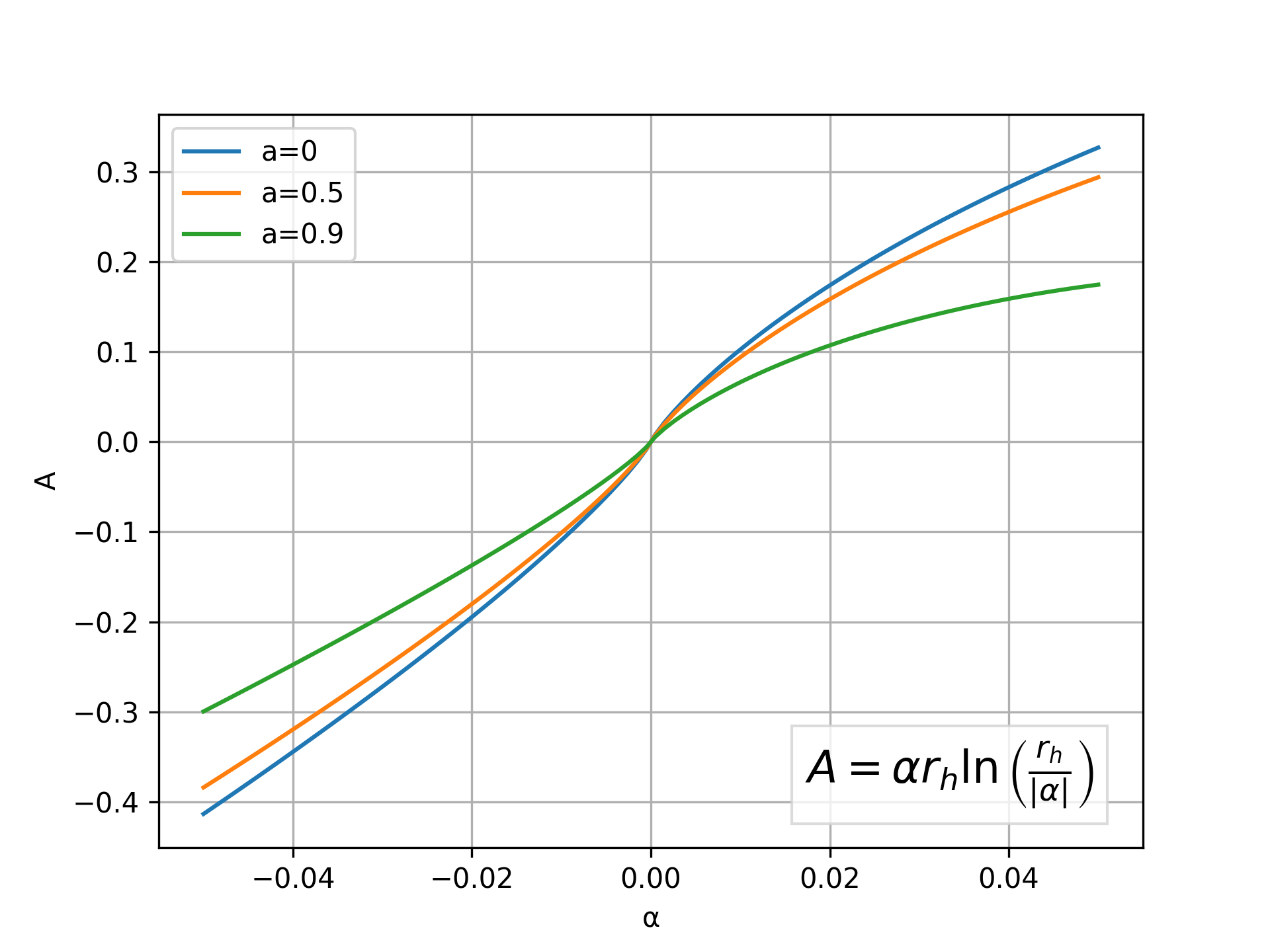}% Here is how to import EPS art
\caption{This is a plot showing the relationship between the parameter $A$ and the strength parameter $\alpha$ of the perfect fluid DM, where the black hole mass $M=1$, the spin parameter $a=0,0.5,0.9$.}
\label{a}
\end{figure}

By referring to Figures 1 and 2, along with expression $A=\alpha\ r_hln{(}\frac{r_h}{\left|\alpha\right|})$, we can derive the following conclusion. When it is $\alpha=\pm r_{h}$ or $\alpha=0$, the parameter $A=0$; when it is $\alpha \in (k_1,-r_{h})\cup(0, r_{h})$, the parameter $A>0$; when it is $\alpha \in (-r_{h},0)\cup(r_{h},k_2)$, the parameter $A<0$. ( $k_1$ and $k_2$ are functions of the black hole spin parameter, satisfying expressions $M^{2} =a^{2} +k_{1}r_{h}\ln(\frac{r_{h} }{-k_{1} }) $ and $M^{2} =a^{2} +k_{2}r_{h}\ln(\frac{r_{h} }{k_{2} }) $. )

So the event horizon $r_h$ can be expressed as
\begin{equation}\label{5}
\frac{r_h}{M}=1\pm\sqrt{1-\frac{a^2+A}{M^2}}.
\end{equation}
The plus sign corresponds to the event horizon, while the minus sign corresponds to the inner horizon. It is evident that the values of the horizons strongly depend on the black hole spin parameter $a$ and the strength parameter $\alpha$. For $a^2+A\le M^2$, the metric describes a black hole, while for $a^2+A>M^2$, the metric describes a rotating spacetime without the event horizon.

The surface area of the black hole’s event horizon is
\begin{equation}\label{6}
A_{H}=4\pi \left ( r_{h}^{2}+a^{2} \right),
\end{equation}
the Hawking temperature associated with the event horizon
\begin{equation}\label{7}
T_H=\frac{r_h^2-a^2}{4\pi r_h\left(r_h^2+a^2\right)},
\end{equation}
the metric described by equation \eqref{1} corresponds to a rotating black hole, with an angular velocity of
\begin{equation}\label{8}
\mathrm{\Omega}_H=\frac{a}{r_h^2+a^2}.
\end{equation}
\section{Verifying the weak cosmic censorship conjecture through a test particle incidence}
In this section, the focus is primarily on exploring the possibility of disrupting the event horizon of a Kerr-like black hole with perfect fluid dark matter by a test particle incident into the black hole. From equation \eqref{5}, we can deduce that when $a^2+A\le M^2$, the spacetime possesses an event horizon; while when $a^2+A>M^2$, the spacetime lacks an event horizon. Hence, the conditions for disrupting the event horizon become
\begin{equation}\label{9}
J>M^2\sqrt{1-\frac{A}{M^2}}=M^2\omega_0,
\end{equation}
there is
\begin{equation}\label{10}
\omega_0=\sqrt{1-\frac{A}{M^2}}.
\end{equation}

To obtain the internal structure of the event horizon of a Kerr-like black hole with perfect fluid dark matter, it is sufficient to inject a test particle or a scalar field with a high angular momentum into the black hole. This allows the Kerr-like black hole to acquire energy from the test particle or scalar field, forming a composite system that disrupts the event horizon. Therefore, the conditions for this composite system to fulfill are given as $J^\prime>M^{\prime2}\omega_0$.

In the spacetime of a Kerr-like black hole with perfect fluid dark matter, if a test particle with mass $M$ is thrown, its motion trajectory can be described by a geodesic. By choosing the affine parameter proper time as the parameter to describe the motion of the test particle, the equation of motion for the test particle can be obtained as given below
\begin{equation}\label{11}
\frac{d^2x^\mu}{d\tau^2}+\mathrm{\Gamma}_{\alpha\beta}^\mu\frac{dx^\alpha}{d\tau}\frac{dx^\beta}{d\tau}=0,
\end{equation}
using the Euler-Lagrange equations, the geodesic equation for the test particle can be obtained from its Lagrangian. In this case, the expression for the Lagrangian describing the particle is
\begin{equation}\label{12}
L=\frac{1}{2}mg_{\mu\nu}\frac{dx^\mu}{d\tau}\frac{dx^\nu}{d\tau}
=\frac{1}{2}mg_{\mu\nu}{\dot{x}}^\mu{\dot{x}}^\nu.
\end{equation}

When slowly incident from infinity onto the equatorial plane, it can be visually observed that the test particle lacks any velocity component in the $\theta$ direction. As a result, its motion before crossing the event horizon will remain confined to the equatorial plane. Therefore, the momentum $P_\theta$ in the $\theta$ direction of the test particle will be zero
\begin{equation}\label{13}
P_\theta=\frac{\partial L}{\partial\dot{\theta}}=mg_{22}\dot{\theta}=0,
\end{equation}
analyzing the motion of the test particle, we find that the energy $\delta E$ and angular momentum  $\delta J$ can be expressed as
\begin{equation}\label{14}
\delta E=-P_t=-\frac{\partial L}{\partial \dot{t}}=-mg_{0\nu} \dot{x}^\nu,
\end{equation}
\begin{equation}\label{15}
\delta J=P_\varphi=\frac{\partial L}{\partial\dot{\varphi}}=mg_{3\nu}\dot{x}^\nu.
\end{equation}

After the black hole captures a test particle with energy $\delta E$ and angular momentum $\delta J$, the energy and angular momentum of the black hole will change according to the following form
\begin{equation}\label{16}
M\rightarrow\ M^\prime=M+\delta E,
\end{equation}
\begin{equation}\label{17}
J\rightarrow\ J^\prime=J+\delta J.
\end{equation}

In this section, we discuss whether the event horizon of a black hole is disrupted when a test particle passes through it. First, we deduce the conditions under which a test particle can enter a Kerr-like black hole, namely energy $\delta E$ and angular momentum $\delta J$. Then, we determine the energy or angular momentum requirements for the disruption of the event horizon based on our calculations for the Kerr-like black hole. As long as the test particle can enter the inner region of the event horizon and the absorbed energy and angular momentum by the Kerr-like black hole satisfy the conditions for event horizon disruption, the internal structure of the event horizon can be revealed to distant observers.

For a test particle with a mass of $m$, the magnitude of its four-velocity is described by a timelike length of $\left|\vec{v}\right|<c$, and its direction can be described by a unit vector $\vec{\mu}$
\begin{equation}\label{18}
U^\mu\ U_\mu=g_{\mu\nu}\frac{dx^\mu}{d\tau}\frac{dx^\nu}{d\tau}=\frac{1}{m^2}g^{\mu\nu}P_\mu\ P_\nu=-1,
\end{equation}
substituting equations \eqref{14} and \eqref{15} into the above expression, we obtain
\begin{equation}\label{19}
g^{00}\delta E^2+g^{11}P_r^2+g^{33}\delta J^2-2g^{03}\delta E\delta J=-m^2,
\end{equation}
taking $\delta E$ as the unknown variable, the above expression is a quadratic equation. By applying the quadratic formula, we can obtain
\begin{equation}\label{20}
\delta E=\frac{g^{03}}{g^{00}}\delta J\pm\frac{1}{g^{00}}\sqrt{(g^{03})^2\delta J^2-g^{00}(g^{33}\delta J^2+g^{11}P_r^2+m^2)}.
\end{equation}

Since the spacetime outside the event horizon is regular, the geodesic motion of the test particle outside the event horizon is timelike and directed towards the future. Therefore, we have
\begin{equation}\label{21}
\frac{dt}{d\tau}>0,
\end{equation}
expanding equations \eqref{14} and \eqref{15}, we obtain
\begin{equation}\label{22}
mg_{00}\dot{t}+mg_{03}\dot{\varphi}=-\delta E,
\end{equation}
\begin{equation}\label{23}
mg_{30}\dot{t}+mg_{33}\dot{\varphi}=\delta J.
\end{equation}
If we consider $\dot{t}$ and $\dot{\varphi}$ as unknown variables, the above equation can be viewed as a system of linear equations. By solving it, we obtain
\begin{equation}\label{24}
\dot{t}=\frac{dt}{d\tau}=-\frac{g_{33}\delta E+g_{03}\delta J}{g_{00}g_{33}-g_{03}^2}.
\end{equation}
By using equations \eqref{21} and \eqref{24}, we can obtain
\begin{equation}\label{25}
\delta E\textgreater-\frac{g_{03}}{g_{33}}\delta J,
\end{equation}
so, the value that satisfies the condition \eqref{25} in equation \eqref{20} is, which corresponds to the energy of the test particle as $\delta E$
\begin{equation}\label{26}
\delta E=\frac{g^{03}}{g^{00}}\delta J-\frac{1}{g^{00}}\sqrt{(g^{03})^2\delta J^2-g^{00}(g^{33}\delta J^2+g^{11}P_r^2+m^2)}.
\end{equation}
By using equation \eqref{24}, we can obtain the upper limit for the value of the test particle $\delta J$
\begin{equation}\label{27}
g_{33}\delta E+g_{03}\delta J>0,
\end{equation}
\begin{equation}\label{28}
\delta J<\frac{r_{h}^{2}+a^2}{a}\delta E =\frac{1}{\mathrm{\Omega}_H}\delta E=\delta J_{max}.
\end{equation}

Furthermore, in order to disrupt the event horizon of a Kerr-like black hole with perfect fluid DM, it can be determined through analyzing the conditions for event horizon disruption of Kerr-like black hole that the angular momentum $\delta J$ of the test particle must be significantly greater than the energy $\delta E$. In other words, the angular momentum $\delta J$ of the test particle should be maximized as much as possible. Therefore, it is also necessary to satisfy the following condition: there should be a lower limit for the value of $\delta J$.
\begin{equation}\label{29}
J+\delta J>(M+\delta E)^2\omega_0,
\end{equation}
\begin{equation}\label{30}
\delta J>\delta J_{min}=\omega _{0}\delta E^2+2\omega _{0}M\delta E+(M^2\omega _{0}-J).
\end{equation}

Therefore, if the test particle simultaneously satisfies both conditions \eqref{28} and \eqref{30}, the event horizon of a Kerr-like black hole with perfect fluid DM will be disrupted, enabling observers at infinity to observe the internal structure of the event horizon. 

For a Kerr-like black hole with perfect fluid DM, if the initial state of the black hole is extremal, it follows that $a^2+A=M^2$. In this scenario, the event horizon of this extremal case can be expressed as
\begin{equation}\label{31}
r_h=M.
\end{equation}

The angular velocity $\mathrm{\Omega}_H$ of this extremal case can be simplified as
\begin{equation}\label{32}
\mathrm{\Omega}_H=\frac{M\omega_0}{2M^2\omega_0^2+A}.
\end{equation}
If we only consider the first-order of energy $\delta E$, then the event horizon of this extremal Kerr-like black hole with perfect fluid DM can only be disrupted if both of the following conditions are satisfied
\begin{equation}\label{33}
\delta J<\delta J_{max}=\frac{1}{\mathrm{\Omega}_H}\delta E=2M\omega _{0}\delta E+\frac{A}{M\omega _{0}}\delta E,
\end{equation}
\begin{equation}\label{34}
\delta J>\delta J_{min}=2M\omega _{0}\delta E.
\end{equation}

Clearly, for an extremal Kerr-like black hole with perfect fluid DM, whether the angular momentum $\delta J$ and energy $\delta E$ of the test particle can simultaneously satisfy conditions \eqref{33} and \eqref{34} depends on the value of parameter $A$, i.e.the strength parameter $\alpha$ of the perfect fluid DM. When $\alpha \in \left (-r_{h} , 0\right ) \cup \left ( r_{h} ,k_2  \right )$ i.e.$(A<0)$, the two equations cannot be satisfied simultaneously. When $\alpha\in\left(k_1,-r_h\right)\cup\left(0,r_h\right)$ i.e.$(A>0)$, the two equations can be simultaneously satisfied, and in this case, the event horizon of the Kerr-like black hole with perfect fluid DM can be disrupted. It is worth noting that when the strength parameter $\alpha=0$ i.e.$A=0$ of the perfect fluid DM , the event horizon of the Kerr black hole cannot be disrupted. At this point, the Kerr black hole becomes an extremal Kerr black hole, which is consistent with the case where the event horizon of an extremal Kerr black hole cannot be disrupted.

If we need to consider the second-order term $\delta E^2$ of the test particle’s energy $\delta E$, it will only increase the lower limit $\delta J_{min}$ of the test particle, making the values more precise. 

For a Kerr-like black hole with perfect fluid DM, if the initial state of the black hole is near-extremal, and the energy $\delta E$ of the test particle is taken at the first-order, the conditions for the test particle to enter the black hole and disrupt its event horizon can be obtained
\begin{equation}\label{35}
\delta J<\delta J_{max}=\frac{1}{\Omega_{H}}\delta E=\frac{r_{h}^{2}+a^2}{a} \delta E,
\end{equation}
\begin{equation}\label{36}
\delta J>\delta J_{min}=2M\omega_{0}\delta E+(M^2\omega_{0}-J).
\end{equation}

To describe the degree of approaching the extremal condition, a dimensionless decimal $\varepsilon$ can be used to depict the near-extremal scenario, i.e.
\begin{equation}\label{37}
\frac{a^2+A}{M^2}=1-\varepsilon^2.
\end{equation}
It can be observed that when parameter $\varepsilon$ approaches zero, i.e.$\varepsilon\rightarrow0$, it describes the near-extremal Kerr-like black hole. And when parameter $\varepsilon=0$ is satisfied, the Kerr-like black holes becomes the extremal case. 

In this context, $M^2\omega_{0}-J$ is a second-order small quantity, but since we are only considering the first-order scenario, this second-order small quantity is not taken into account. Therefore, it is sufficient for the energy $\delta E$ and angular momentum $\delta J$ of the test particle to simultaneously satisfy equations \eqref{35} and \eqref{36}, i.e.
\begin{equation}\label{38}
\delta J_{max}>\delta J_{min},
\end{equation}
this is equivalent to
\begin{equation}\label{39}
\frac{1}{\Omega_{H}}-2\omega_{0}M>0.
\end{equation}
Based on equations \eqref{5}, \eqref{10}, and \eqref{37}, we obtain
\begin{equation}\label{40}
\begin{split}
&\frac{1}{\Omega}_{H}-2\omega_{0}M=\frac{r_{h}^{2}+a^2}{a}-2\omega_{0}M
\\&=\frac{A+\left(2M^2\omega_0^2+2A\right)\varepsilon+\left(M^2\omega_0^2+A\right)\varepsilon^2-2M^2\omega_0^2O\left(\varepsilon^4\right)}{\sqrt{M^2\omega_0^2-\left(M^2\omega_0^2+A\right)\varepsilon^2}},
\end{split}
\end{equation}
in equation \eqref{40}, $O\left(\varepsilon^4\right)$ is a higher-order infinitesimal, and whether the equation is greater than zero or less than zero depends on the value of parameter $A$, which is the strength parameter $\alpha$ of the perfect fluid DM.

When $\alpha\in\left(-r_h,0\right)\cup\left(r_h,k_2\right)$ i.e.$(A<0)$ 
\begin{equation}\label{41}
\frac{1}{\Omega_{H}}-2\omega_{0}M<0,
\end{equation}
at this point, the event horizon of the near-extremal Kerr-like black hole with perfect fluid DM cannot be disrupted, and this near-extremal scenario does not violate the cosmic censorship conjecture.
\\When $\alpha\in\left(k_1,-r_h\right]\cup\left[0,r_h\right]$ i.e.$(A\ge 0)$ 
\begin{equation}\label{42}
\frac{1}{\Omega_{H}}-2\omega_{0}M>0,
\end{equation}
the event horizon of the near-extremal Kerr-like black hole with perfect fluid DM can be disrupted, and this near-extremal scenario violates the cosmic censorship conjecture. It is worth noting that when the strength parameter $\alpha=0$, i.e.$A=0$, the metric degenerates into a near-extremal Kerr black hole. From the above equation, it can be seen that in this case, the event horizon can be destroyed, which is consistent with previous research showing that a near-extremal Kerr black hole can be disrupted by a test particle, revealing the singularity\cite{jacobson2009}.

Based on the analysis above, for incoming a test particle to disrupt the event horizon of a Kerr-like black hole with perfect fluid DM, the ability to disrupt the event horizon depends on $A$, where $A$ is a function of $\alpha$. For the near-extremal scenario of this Kerr-like black hole, when $\alpha\in\left(-r_h,0\right)\cup\left(r_h,k_2\right)$ i.e.$(A<0)$, the event horizon of the black hole cannot be disrupted, adhering to the cosmic censorship conjecture. However, when $\alpha\in\left(k_1,-r_h\right]\cup\left[0,r_h\right]$ i.e.$(A\ge 0)$, the event horizon of the black hole can be disrupted, thus violating the cosmic censorship conjecture.
\section{Verifying the weak cosmic censorship conjecture through a scalar field incidence}
Another way to disrupt the event horizon of a Kerr-like black hole with perfect fluid DM is by throwing a scalar field with high angular momentum into the black hole, which would result in the destruction of the event horizon. Semiz\cite{semiz2011} initially proposed a thought experiment involving the introduction of a classical field into extremal and near-extremal cases of black holes to destroy their event horizons. Research has shown that in this situation, the event horizon of the black hole cannot be destroyed\cite{semiz2011}. Others have further developed this method\cite{duztas2013,semiz2015,gwak2018,Dzta2019KerrNewmanBH}.

In this section, the focus is primarily on introducing a scalar field with high angular momentum into the interior of a Kerr-like black hole and discussing the potential for event horizon disruption in extremal and near-extremal scenarios of the Kerr-like black hole.
\subsection{The scattering of a massive classical scalar field}
In this subsection, we investigate the mass scattering of a classical scalar field with a Kerr-like black hole that possess perfect fluid DM. The scalar field $\psi$ has a mass of $\mu$. The equation for the scalar field $\psi$ can be derived from the Klein-Gordon equation.
\begin{equation}\label{43}
\mathrm{\nabla}_\nu\mathrm{\nabla}^\nu\mathrm{\Psi}-\mu^2\mathrm{\Psi}=0,
\end{equation}
in the above equation, $\mathrm{\nabla}_\nu$ represents the covariant derivative. The equation can also be written as follows
\begin{equation}\label{44}
\frac{1}{\sqrt{-g}}\partial_\mu\left(\sqrt{-g}g^{\mu\nu}\partial_\nu\mathrm{\Psi}\right)-\mu^2\mathrm{\Psi}=0.
\end{equation}
By substituting the metric \eqref{1} of the Kerr-like black hole with perfect fluid DM into \eqref{44}, the following equation can be obtained, here we set $\sigma=2Mr-\alpha\ rln{(}\frac{r}{\left|\alpha\right|})$
\begin{equation}\label{45}
\begin{split}
&-\frac{(r^2+a^2)^2-a^2\mathrm{\Delta}_r{sin}^2{\theta}}{\mathrm{\Delta}_r\mathrm{\Sigma}^2}\frac{\partial^2\mathrm{\Psi}}{\partial\ t^2}-\frac{2a\sigma}{\mathrm{\Delta}_r\mathrm{\Sigma}^2}\frac{\partial^2\mathrm{\Psi}}{\partial\ t\partial\varphi}\\&+\frac{1}{\mathrm{\Sigma}^2}\frac{\partial}{\partial\ r}\left(\mathrm{\Delta}_r\frac{\partial\Psi}{\partial\ r}\right)+\frac{1}{\mathrm{\Sigma}^2sin{\theta}}\frac{\partial}{\partial\theta}\left(sin{\theta}\frac{\partial\Psi}{\partial\theta}\right)
\\&+\frac{\mathrm{\Delta}_r-a^2{sin}^2{\theta}}{\mathrm{\Delta}_r\mathrm{\Sigma}^2sin{\theta}}\frac{\partial^2\mathrm{\Psi}}{\partial\varphi^2}-\mu^2\mathrm{\Psi}=0.
\end{split}
\end{equation}
Equation \eqref{45} can be separated into variables\cite{brahma2021}. To facilitate obtaining a solution for the above equation, we can employ the method of variable separation by making the following decomposition
\begin{equation}\label{46}
\mathrm{\Psi}\left(t,r,\theta,\varphi\right)=e^{-i\omega t}R\left(r\right)S_{lm}\left(\theta\right)e^{im\varphi},
\end{equation}
in the above equation, $S_{lm}\left(\theta\right)$ represents the spherical function, where $l$ and $m$ are constant variables representing the separation of variables, denoted as $l,m=1,2,3...$. By substituting equation \eqref{46} into equation \eqref{45} and simplifying, we obtain
\begin{equation}\label{47}
\begin{split}
&\frac{(r^2+a^2)^2-a^2\mathrm{\Delta}_r{sin}^2{\theta}}{\mathrm{\Delta}_r}\omega^2R\left(r\right)S_{lm}\left(\theta\right)-\\&\frac{2a\sigma}{\mathrm{\Delta}_r}m\omega\ R\left(r\right)S_{lm}\left(\theta\right)+S_{lm}\left(\theta\right)\frac{d}{dr}\left(\mathrm{\Delta}_r\frac{dR\left(r\right)}{dr}\right)+\\&\frac{1}{sin{\theta}}R\left(r\right)\frac{d}{d\theta}\left(sin{\theta}\frac{dS_{lm}\left(\theta\right)}{d\theta}\right)+\\&\frac{a^2{sin}^2{\theta}-\mathrm{\Delta}_r}{\mathrm{\Delta}_rsin{\theta}}m^2R\left(r\right)S_{lm}\left(\theta\right)-\\&\left(r^2+a^2{cos}^2{\theta}\right)\mu^2R\left(r\right)S_{lm}\left(\theta\right)=0.
\end{split}
\end{equation}
By separating variables in \eqref{47}, we can obtain the angular part of the motion equation for the scalar field as
\begin{equation}\label{48}
\begin{split}
&\frac{1}{sin{\theta}}\frac{d}{d\theta}\left[sin{\theta}\frac{dS_{lm}\left(\theta\right)}{d\theta}\right]-\\&\left[a^2\omega^2{sin}^2{\theta}+\frac{m^2}{{sin}^2{\theta}}+\mu^2a^2{cos}^2{\theta}-\lambda_{lm}\right]S_{lm}=0,
\end{split}
\end{equation}
and the radial part of the motion equation for the scalar field is
\begin{equation}\label{49}
\begin{split}
&\frac{d}{dr}\left(\mathrm{\Delta}_r\frac{dR}{dr}\right)+\\&\left[\frac{(r^2+a^2)^2}{\mathrm{\Delta}_r}\omega^2-\frac{2a\sigma}{\mathrm{\Delta}_r}m\omega+\frac{m^2a^2}{\mathrm{\Delta}_r}-\mu^2r^2-\lambda_{lm}\right]R\left(r\right)=0.
\end{split}
\end{equation}

In the motion equation for the scalar field, the solution of the angular part \eqref{48} is a spherical function with an eigenvalue of $\lambda_{lm}$\cite{seidel1989}. When calculating the stress-energy tensor later, the integration is performed over the entire event horizon surface. According to the normalization property of spherical functions, the integral over the event horizon surface is equal to one. Therefore, the specific expression of the spherical function is not separately solved here. Instead, the focus is on solving the radial equation.

To facilitate the solution of the radial equation, the tortoise coordinate $r_\ast$ is introduced
\begin{equation}\label{50}
\frac{dr}{dr_\ast}=\frac{\mathrm{\Delta}_r}{r^2+a^2}.
\end{equation}

After introducing the tortoise coordinate, the event horizon is pushed to infinity, and the tortoise coordinate takes a value of $\left(-\infty,+\infty\right)$ in this case. Therefore, the radial coordinate covers the entire region from the event horizon outward in the spacetime. By substituting the tortoise coordinate from equation \eqref{50} into the radial component equation of the scalar field, namely equation \eqref{49}, the radial equation simplifies to the following form
\begin{equation}\label{51}
\begin{split}
\frac{\mathrm{\Delta}_r}{(r^2+a^2)^2}\frac{d}{dr}\left(r^2+a^2\right)\frac{dR}{dr_\ast}+\frac{d^2R}{dr_\ast^2}+\left[\left(\omega-\frac{ma}{r^2+a^2}\right)^2+\right.\\
\left.\frac{\mathrm{\Delta}_r}{(r^2+a^2)^2}2am\omega-\frac{\mathrm{\Delta}_r}{(r^2+a^2)^2}\left(\mu^2r^2+\lambda_{lm}\right)\right]R\left(r\right)=0.
\end{split}
\end{equation}

During the entire process of the scalar field incident on this spacetime, we pay closer attention to the situation near the event horizon, especially in terms of energy flux and angular momentum flux in this region, namely near point $r\cong r_h$. In this vicinity, the equation for solving the event horizon, denoted as $\mathrm{\Delta}_r\cong0$, is taken into account. Therefore, the radial equation for the scalar field in equation \eqref{51} can be approximated and simplified as follows
\begin{equation}\label{52}
\frac{d^2R}{dr_\ast^2}+\left(\omega-\frac{ma}{r^2+a^2}\right)^2R=0.
\end{equation}
Due to the fact that the angular velocity of this Kerr spacetime at the event horizon is
\begin{equation}\label{53}
\mathrm{\Omega}_H=\frac{a}{r_h^2+a^2}.
\end{equation}
By substituting \eqref{53} into the simplified radial equation \eqref{52}, it can be further simplified to
\begin{equation}\label{54}
\frac{d^2R}{dr_\ast^2}+\left(\omega-m\mathrm{\Omega}_H\right)^2R=0,
\end{equation}
the above equation is a second-order homogeneous partial differential equation, so the solution to the equation is
\begin{equation}\label{55}
R\left(r\right)=exp{\left[\pm\ i\left(\omega-m\mathrm{\Omega}_H\right)r_\ast\right]},
\end{equation}
in the above equation, the positive and negative solutions represent the outgoing wave and the ingoing wave, respectively. When considering the scalar field incident on the Kerr spacetime, this spacetime absorbs energy and angular momentum. Therefore, in this case, the negative solution is more consistent with the physical reality. Thus, the radial solution of the scalar field equation is
\begin{equation}\label{56}
R\left(r\right)=exp{\left[-i\left(\omega-m\mathrm{\Omega}_H\right)r_\ast\right]}.
\end{equation}
Substituting the radial solution \eqref{56} into the solution of the scalar field equation \eqref{46}, we obtain the following expression
\begin{equation}\label{57}
\mathrm{\Psi}\left(t,r,\theta,\varphi\right)=exp{\left[-i\left(\omega-m\mathrm{\Omega}_H\right)\right]}e^{-i\omega t}S_{lm}\left(\theta\right)e^{im\varphi}.
\end{equation}

The above expression represents the solution of the scalar field near the event horizon when a massive scalar field is incident on the Kerr spacetime. By obtaining this solution, it is possible to calculate the flux of energy and angular momentum absorbed by the Kerr spacetime from the scalar field after scattering. This can be used to discuss whether the event horizon of the Kerr spacetime can be destroyed after absorbing energy.

The next step is to primarily study the calculation of the energy flux and angular momentum flux brought about by the scattering of the scalar field from the event horizon of the Kerr spacetime. Here, we assume that the scalar field has a mode of $\left(l,m\right)$, and it is incident upon the Kerr spacetime. During this process, a portion of the scalar field’s energy is absorbed by the spacetime, while the rest is reflected back. The main focus and concern lie in the angular momentum and energy absorbed by the Kerr spacetime, exploring whether the destruction of the event horizon of the spacetime can occur after absorbing angular momentum and energy.

The energy-momentum tensor $T_{\mu\nu}$ of the scalar field with a mass of $\mu$ can be expressed in the following form
\begin{equation}\label{58}
T_{\mu\nu}=\partial_\mu\mathrm{\Psi}\partial_\nu\mathrm{\Psi}^\ast-\frac{1}{2}g_{\mu\nu}\left(\partial_\alpha\mathrm{\Psi}\partial^\alpha\mathrm{\Psi}^\ast+\mu^2\mathrm{\Psi}^\ast\mathrm{\Psi}\right),
\end{equation}
substituting the metric of the Kerr-like black hole into equation \eqref{58}, we obtain
\begin{equation}\label{59}
T_t^r=\frac{r^2+a^2}{\mathrm{\Sigma}^2}\omega\left(\omega-m\mathrm{\Omega}_H\right)S_{lm}\left(\theta\right)e^{im\varphi}{S^\ast}_{l^,m^,}\left(\theta\right)e^{-im\varphi},
\end{equation}
\begin{equation}\label{60}
T_\varphi^r=\frac{r^2+a^2}{\mathrm{\Sigma}^2}m\left(\omega-m\mathrm{\Omega}_H\right)S_{\mathrm{lm}}\left(\theta\right)e^{im\varphi}{S^\ast}_{l^,m^,}\left(\theta\right)e^{-im\varphi}.
\end{equation}
From the above equation, the energy flux passing through the event horizon of the Kerr spacetime can be calculated as
\begin{equation}\label{61}
\frac{dE}{dt}=\omega\left(\omega-m\mathrm{\Omega}_H\right)\left(r^2+a^2\right),
\end{equation}
the angular momentum flux passing through the event horizon of the Kerr spacetime is
\begin{equation}\label{62}
\frac{dJ}{dt}=m\left(\omega-m\mathrm{\Omega}_H\right)\left(r^2+a^2\right).
\end{equation}
In equations \eqref{61} and \eqref{62}, the integration involves the normalization of spherical harmonics.

By observing the above two equations, it can be intuitively understood that when $\omega>m\mathrm{\Omega}_H$, the angular momentum flux and energy flux passing through the event horizon are positive. This indicates that when the massive scalar field scatters into the Kerr spacetime, the Kerr black hole absorbs energy and angular momentum from the scalar field. When $\omega<m\mathrm{\Omega}_H$, the energy flux and angular momentum flux passing through the event horizon are negative. This signifies that the Kerr black hole does not absorb energy and angular momentum from the scalar field, but rather the scalar field carries away energy and angular momentum from the Kerr black hole during the scattering process, known as the famous black hole superradiance\cite{brito2015}.

If we only consider a very short time interval of $\mathrm{d}t$, the amount of energy absorbed by the Kerr-like black hole from the scalar field is
\begin{equation}\label{63}
dE=\omega\left(\omega-m\mathrm{\Omega}_H\right)\left(r^2+a^2\right)\mathrm{\mathrm{dt}},
\end{equation}
the absorbed angular momentum is
\begin{equation}\label{64}
dJ=m\left(\omega-m\mathrm{\Omega}_H\right)\left(r^2+a^2\right)\mathrm{\mathrm{dt}}.
\end{equation}

With the variation in the absorbed angular momentum and energy from the scalar field by the Kerr spacetime, it is possible to analyze whether the event horizon of the Kerr spacetime can be broken by a scalar field with high angular momentum in extremal and near-extremal cases. If it can be broken, it would expose the internal structure of the event horizon of the Kerr black hole to external observers.
\subsection{Disruption of the event horizon of a Kerr-like black hole through the monochromatic classical scalar field incidence}
In this section, we verify whether the event horizon of the Kerr-like black hole with perfect fluid dark matter (DM) can be broken by an incident classical scalar field with a frequency of $\omega$ and an angular quantum number of $m$. We focus on a small time interval of $\mathrm{d}t$ during the scattering process. For longer scattering times, we can employ a differential approach by dividing it into numerous infinitesimal time intervals of $\mathrm{d}t$, thus allowing us to analyze the outcome by examining any specific time interval of $\mathrm{d}t$. Such an approach simplifies the treatment of scattering problems and enhances our understanding of the physical processes and laws involved in scattering. Importantly, studying each time interval of $\mathrm{d}t$ provides a more detailed and comprehensive scattering information, enabling further investigations into the internal mechanisms involved. During the scattering process, we contemplate a composite system where a Kerr-like black hole with perfect fluid DM, having a mass of $M$ and an angular momentum of $J$, absorbs the energy and angular momentum of the incident scalar field, transforming into a composite system with a mass of $M^\prime$ and an angular momentum of $J^\prime$. In order to validate the cosmic censorship conjecture, it is necessary to consider the sign of the parameter $M^{\prime2}\omega_0-J^\prime$ in the composite system. If $M^{\prime2}\omega_0-J^\prime\geq0$, then the event horizon of the Kerr-like black hole exists, thus not violating the cosmic censorship conjecture. However, if $M^{\prime2}\omega_0-J^\prime<0$, then the event horizon of the Kerr-like black hole does not exist, thereby violating the cosmic censorship conjecture.

During the scattering process, the focus is primarily on the scattering events within a small time interval of $\mathrm{d}t$. After the composite system absorbs the energy and angular momentum of the incident scalar field, the state of the composite system is
\begin{equation}\label{65}
M^{\prime2}\omega_0-J^\prime=\left(M^2\omega_0-J\right)+2M\omega_0dE-dJ.
\end{equation}
Substituting equations \eqref{63} and \eqref{64} into equation \eqref{65}, we obtain
\begin{equation}\label{66}
\begin{split}
&M^{\prime2}\omega_0-J^\prime=\left(M^2\omega_0-J\right)+\\&2M\omega_0m^2\left(\frac{\omega}{m}-\frac{1}{2M\omega_0}\right)\left(\frac{\omega}{m}-\mathrm{\Omega}_H\right)\left(r^2+a^2\right)dt.
\end{split}
\end{equation}

When the initial state of the Kerr-like black hole with perfect fluid DM is in the extremal case, we have $J=M^2\omega_0$. Therefore, equation \eqref{66} can be rewritten as
\begin{equation}\label{67}
\begin{split}
&M^{\prime2}\omega_0-J^\prime=\\&2M\omega_0m^2\left(\frac{\omega}{m}-\frac{1}{2M\omega_0}\right)\left(\frac{\omega}{m}-\mathrm{\Omega}_H\right)\left(r^2+a^2\right)dt.
\end{split}
\end{equation}
The angular velocity $\mathrm{\Omega}_H$ of the Kerr-like black hole in the extremal case can be simplified as 
\begin{equation}\label{68}
\mathrm{\Omega}_H=\frac{a}{r_h^2+a^2}=\frac{M\omega_0}{2M^2\omega_0^2+A}.
\end{equation}

If the perfect fluid does not exist, meaning $\alpha=0$, the metric described by equation \eqref{1} will become the metric of a Kerr black hole. In the extremal case, the angular velocity of the Kerr black hole with perfect fluid DM is influenced by the strength parameter $\alpha$, resulting in a difference from the angular velocity of the standard Kerr black hole. This difference plays a crucial role in the scattering process of the scalar field.

We incident the following mode of scalar field into the extremal case of the Kerr black hole
\begin{equation}\label{69}
\frac{\omega}{m}=\frac{1}{2}\left(\frac{1}{2M\omega_0}+\mathrm{\Omega}_H\right),
\end{equation}
substituting equation \eqref{69} into the composite system of Kerr black hole, which absorbs the energy and angular momentum of the scalar field, we obtain by substituting into equation \eqref{67}
\begin{equation}\label{70}
M^{\prime2}\omega_0-J^\prime=-\frac{1}{2}M\omega_0m^2(\mathrm{\Omega}_H-\frac{1}{2M\omega_0})^2(r^2+a^2)dt.
\end{equation}
For the extremal case of the Kerr black hole, we can obtain from equation \eqref{70}
\begin{equation}\label{71}
M^{\prime2}\omega_0-J^\prime\le0.
\end{equation}

When the PFDM does not exist, i.e.$\alpha=0$ (parameter $A=0$), the equation above holds true, which corresponds to the Kerr black hole. In this case, the event horizon of the black hole cannot be destroyed\cite{yang2020}, aligning with the predictions of general relativity that a massive scalar field cannot disrupt the event horizon of an extremal Kerr black hole. However, interestingly, through the analysis of the equation above, we find that when parameter $\alpha\neq0 $ and $\alpha\neq\pm\ r_h$ i.e.$(A\neq0)$, $M^{\prime2}\omega_0-J^\prime<0$ holds true. That is, in this scenario, the event horizon of the Kerr-like black hole can be destroyed.

By analyzing equations \eqref{67} and \eqref{68} in combination, we can observe that there exists a range of scalar field modes that can potentially disrupt the event horizon.
\\When $\alpha\in\left(k_1,-r_h\right)\cup\left(0,r_h\right)$ i.e.$(A>0)$  
\begin{equation}\label{72}
\mathrm{\Omega}_H=\frac{a}{r_h^2+a^2}<\frac{1}{2M\omega_0},
\end{equation}
in this case, the range of scalar field modes is
\begin{equation}\label{73}
\mathrm{\Omega}_H<\frac{\omega}{m}<\frac{1}{2M\omega_0}.
\end{equation}
\\When $\alpha\in\left(-r_h,0\right)\cup\left(r_h,k_2\right)$ i.e.$(A<0)$ 
\begin{equation}\label{74}
\mathrm{\Omega}_H=\frac{a}{r_h^2+a^2}>\frac{1}{2M\omega_0},
\end{equation}
in this case, the range of scalar field modes is
\begin{equation}\label{75}
\frac{1}{2M\omega_0}<\frac{\omega}{m}<\mathrm{\Omega}_H.
\end{equation}
In other words, for different values of the strength parameter $\alpha$ i.e. parameter $A$ of the perfect fluid DM, there exist different ranges of scalar field modes that can disrupt the event horizon. Within these ranges, the event horizon can be destroyed.

When the initial state of the Kerr black hole with perfect fluid DM approaches the near-extremal case, i.e.$J\neq\ M^2\omega_0$, but tends to infinity, it can be expressed using the following formula
\begin{equation}\label{76}
\begin{split}
&M^{\prime2}\omega_0-J^\prime=\left(M^2\omega_0-J\right)+\\&2M\omega_0m^2\left(\frac{\omega}{m}-\frac{1}{2M\omega_0}\right)\left(\frac{\omega}{m}-\mathrm{\Omega}_H\right)\left(r^2+a^2\right)dt,
\end{split}
\end{equation}
similarly to the extremal case, considering the mode of the scalar field incident in this near-extremal situation is
\begin{equation}\label{77}
\frac{\omega}{m}=\frac{1}{2}\left(\mathrm{\Omega}_H+\frac{1}{2M\omega_0}\right).
\end{equation}
Therefore, equation \eqref{76} can be expressed in the following form
\begin{equation}\label{78}
\begin{split}
&M^{\prime2}\omega_0-J^\prime=(M^2\omega_0-J)-\\&\frac{1}{8M\omega_0}m^2\mathrm{\Omega}_H^2(\frac{1}{\mathrm{\Omega}_H}-2M\omega_0)^2(r^2+a^2)dt.
\end{split}
\end{equation}

Similar to the previously discussed dimensionless infinitesimal parameter $\varepsilon$ that describes the destruction of the event horizon by a test particle incident on a near-extremal Kerr-like black hole, we now define another dimensionless infinitesimal parameter $\varepsilon$ to describe the deviation of a near-extremal Kerr-like black hole from an extremal Kerr-like black hole situation
\begin{equation}\label{79}
\frac{a^2}{M^2{\omega_0}^2}=1-\varepsilon^2,
\end{equation}
it can be concluded that as parameter $\varepsilon$ decreases, the black hole approaches a near-extremal situation; when parameter $\varepsilon=0$ is reached, the black hole becomes extremal.

Since $\varepsilon$ is a parameter that tends to zero, it can be expanded using a Taylor series. Therefore, equation \eqref{78} can be expressed in the following form
\begin{equation}\label{80}
\begin{split}
&M^{\prime2}\omega_0-J^\prime=[\frac{1}{2}\frac{M^{2}\omega_{0}^{2}+A}{\omega_{0}}\varepsilon^{2}-M^2\omega_{0}O\left(\varepsilon^4\right)]-\\&\frac{1}{8M\omega_0}m^2\mathrm{\Omega}_H^2\times\\&\left(\frac{A+\left(2M^2\omega_0^2+2A\right)\varepsilon+\left(M^2\omega_0^2+A\right)\varepsilon^2-2M^2\omega_0^2O\left(\varepsilon^4\right)}{\sqrt{M^2\omega_0^2-\left(M^2\omega_0^2+A\right)\varepsilon^2}}\right)^2\\&\times\left(r^2+a^2\right)dt.
\end{split}
\end{equation}

Since we are considering a very short time interval $\mathrm{d}t$, both $\varepsilon$ and $\mathrm{d}t$ are first-order small quantities. From the equation, we can see that the first part is at most a second-order small quantity, while the second part is predominantly a first-order small quantity, and the latter part is even larger than the former part. In this case
\\when $\alpha\neq0 $ and $\alpha\neq\pm\ r_h$ i.e.$(A\neq0)$,there exists
\begin{equation}\label{81}
M^{\prime2}\omega_0-J^\prime<0,
\end{equation}
at this moment, the introduction of the scalar field causes the event horizon of the black hole to be disrupted.
\\When parameter $A=0$, which represents the strength parameter $\alpha=0$ or $\alpha=\pm r_h$ of the perfect fluid DM, the first part of the equation is at most a second-order small quantity, while the second part is at most a third-order small quantity. Therefore, we have
\begin{equation}\label{82}
M^{\prime2}\omega_0-J^\prime>0.
\end{equation}
When the strength parameter $\alpha=0$ for the perfect fluid DM is considered, the corresponding near-extremal Kerr-like black hole, as can be deduced from the above equation, reveals that the event horizon of a near-extremal Kerr black hole cannot be disrupted by the scalar field. This is consistent with research in general relativity, which suggests that the event horizon of a near-extremal Kerr black hole cannot be destroyed\cite{gwak2018}.

In general, whether the scalar field can disrupt the event horizon of a near-extremal Kerr black hole depends on the value of the strength parameter $\alpha$ (i.e. parameter $A$) for the perfect fluid DM. Only when parameter $\alpha\neq0 $ and $\alpha\neq\pm\ r_h$ i.e.$(A\neq0)$ is reached can the event horizon of a near-extremal Kerr black hole be disrupted by the scalar field.
\section{Summary and discussion}
The dark matter-black hole system involves the interaction between a black hole immersed in a dark matter background, eventually forming a stable system. This system is considered the most plausible model in the universe, making the investigation of the Weak Cosmic Censorship Conjecture (WCCC) significant. The WCCC is a fundamental hypothesis in black hole physics, although its correctness has not been conclusively proven yet and there is no rigorous mathematical formula to support it. However, studying the WCCC is crucial for delving into the underlying physical nature behind the event horizon of black holes. If the WCCC is violated, it would aid physicists in gaining a deeper understanding of the characteristics of highly curved regions within black holes, ultimately revealing profound laws and phenomena within black hole physics.

In this paper, we investigated the incidence of both a test particle and a scalar field into a Kerr-like black hole with perfect fluid dark matter. We explored whether the event horizon of the black hole is disrupted under extremal and near-extremal conditions, in order to test the validity of the Weak Cosmic Censorship Conjecture (WCCC) for the Kerr-like black hole.

We found that when a test particle is incident into the black hole, under extremal conditions, the event horizon of the black hole remains intact when condition $\alpha\in\left(-r_h,0\right)\cup\left(r_h,k_2\right)$ i.e.$(A<0)$ is met, while it is disrupted when condition $\alpha\in\left(k_1,-r_h\right)\cup\left(0,r_h\right)$ i.e.$(A>0)$ is satisfied. Under near-extremal conditions, the event horizon of the black hole is not destroyed when condition $\alpha\in\left(-r_h,0\right)\cup\left(r_h,k_2\right)$ i.e.$(A<0)$ is met, but it is disrupted when condition $\alpha\in\left(k_1,-r_h\right]\cup\left[0,r_h\right]$ i.e.$(A\ge 0)$ is satisfied. When a classical scalar field is incident into the black hole, under extremal conditions, the event horizon can be disrupted when condition $\alpha\neq0 $ and $\alpha\neq\pm\ r_h$ i.e.$(A\neq0)$ is met for a specified mode of scalar field incidence. However, when the mode of scalar field incidence is not specified, we observed that the range of scalar field patterns that can destroy the black hole event horizon varies depending on the strength parameter $\alpha$ of the perfect fluid dark matter. Under near-extremal conditions, through analysis, we determined that the event horizon of the black hole can be disrupted when condition $\alpha\neq0 $ and $\alpha\neq\pm\ r_h$ i.e.$(A\neq0)$ is satisfied.

Furthermore, we observed that both the incidence of a test particle and a scalar field into this Kerr-like black hole violate the Weak Cosmic Censorship Conjecture, potentially offering insights into the inner regions of black hole event horizons for future studies. However, it is important to note that the dark matter model used in this paper is PFDM, which lacks generality. We hope to explore generalized solutions of gravitational field equations that are suitable for dark matter in future research, and then extend and deepen the study presented in this paper to make it more universally applicable and informative.
\section*{Acknowledgements}
We acknowledge the anonymous referee for a constructive report that has significantly improved this paper. This work was supported by the  Special Natural Science Fund of Guizhou University (Grants
No. X2020068, No. X2022133) and the National Natural Science Foundation of China (Grant No. 12365008).

\nocite{*}
\bibliographystyle{unsrt}
\bibliography{meng}

\end{document}